\documentclass{sadhana}
\usepackage{graphicx}
\usepackage{amsmath, amsfonts, bm}
\usepackage[utf8]{inputenc}
\usepackage{textcomp}
\usepackage{soul}
\usepackage{booktabs}
\usepackage{tabularx}
\usepackage{array}
\usepackage{tikz}
\usetikzlibrary{arrows.meta, shapes.geometric, positioning}
\usepackage{algorithm}
\usepackage{algpseudocode}
\usepackage{subcaption}
\usepackage{balance}
\usepackage{url}
\usepackage[none]{hyphenat} 
\newcommand{\authcite}[2]{\textit{#1}~\cite{#2}}
\usepackage{orcidlink}

\begin{document}


\title{Adaptive Model Predictive Control for Ground Vehicles:\\ Review and Demonstrative Implementation}


\author{
Chetana Gadgil\textsuperscript{1}\,\orcidlink{0009-0009-7725-4715},
Mahendra Singh Tomar\textsuperscript{1},\orcidlink{0000-0003-4147-6032}
}
\affilOne{\textsuperscript{1} Department of Electrical and Electronics Engineering, BITS Pilani Goa\\}


\twocolumn[{

\maketitle

\begin{abstract}
This paper reviews Adaptive Model Predictive Control (AMPC) methods for Autonomous Vehicles (AVs), focusing on control strategies that dynamically adapt to uncertainties and changing conditions in real-time. The critical role of Adaptive Model Predictive Control (AMPC) in addressing the challenges of autonomous vehicle control are discussed. For the scope of this paper, AMPC is defined as a class of Model Predictive Control (MPC) techniques that modify the system model, cost function, constraints, or prediction horizon, based on real-time data. Traditional MPC, while effective for constrained optimization, struggles with model inaccuracies, computational demands, and dynamic environments, necessitating AMPC methods. The review covers existing literature on Gain scheduled MPC, Online Model Estimation MPC, Weight Adaptive MPC, Horizon Adaptive MPC, Learning Based MPC, and Hybrid MPC that combines MPC with other control methods. In addition to the survey, a demonstrative simulation of an adaptive MPC controller is presented that illustrates practical aspects of weight and speed adaptation in trajectory tracking.
\end{abstract}



}]




\markboth{Chetana Gadgil, Mahendra Singh Tomar}{Adaptive MPC: Review and Demonstrative Study}

\section{Introduction}
Autonomy in transport has evolved over time from no automation to High Automation over the years. It is advantageous to define the levels of autonomy for regulatory purposes. SAE International’s J3016 standard \cite{SAEJ3016_202104} defines six levels of driving automation, from Level 0 (no automation) to Level 5 (full automation), as shown in Table \ref{tab:autonomy_levels}. These levels provide a framework for understanding the progression of autonomous control, with higher levels demanding more sophisticated control strategies. The current implementations of autonomy are between level 2 to 4 in the industry and research. Achieving Level 5 automation necessitates further research in advanced control strategies for autonomous vehicles.

\begin{table}[t]
\caption{SAE Levels of Driving Automation (J3016 Standard)}
\label{tab:autonomy_levels}
\centering
\footnotesize
\renewcommand{\arraystretch}{1.2}

\begin{tabularx}{\columnwidth}{l c >{\raggedright\arraybackslash}X}
\hline
\textbf{Level} & \textbf{Name} & \textbf{Description} \\
\hline
\textbf{Level 0} & No Automation & Human driver performs all driving tasks. Vehicle may issue warnings or momentary assistance (e.g., lane departure alert). \\
\textbf{Level 1} & Driver Assistance & Vehicle can assist with either steering or acceleration/deceleration (e.g., Adaptive Cruise Control), but not both simultaneously. Driver remains fully engaged.\\
\textbf{Level 2} & Partial Automation & Vehicle can control both steering and acceleration/deceleration under some conditions. Driver must monitor and be ready to intervene at all times.\\
\textbf{Level 3} & Conditional Automation & Vehicle can perform all driving tasks in specific conditions. Human driver must take over when requested (e.g., traffic jam pilot).\\
\textbf{Level 4} & High Automation & Vehicle performs all driving tasks in defined areas or scenarios (geofenced). No driver attention is required within these conditions.\\
\textbf{Level 5} & Full Automation & Vehicle is capable of all driving tasks under all conditions. No driver required at any time or location.\\
\hline
\end{tabularx}

\end{table}

\begin{figure}[!t]
\centering
\begin{tikzpicture}[node distance=1.5cm, auto]
\tikzstyle{block} = [rectangle, draw, fill=blue!10, 
    text width=6cm, text centered, rounded corners, minimum height=3.5em, font=\small]
\tikzstyle{arrow} = [thick,->,>=stealth]

\node [block] (perception) {
    \textbf{Perception} \\
    Cameras, LiDAR, Radar, \\
    Ultrasonic Sensors, \\
    Computer Vision, \\
    Sensor Fusion
};
\node [block, below=of perception] (localization) {
    \textbf{Localization} \\
    GPS, IMU, \\
    Odometry, \\
    SLAM, \\
    HD Maps
};
\node [block, below=of localization] (path_planning) {
    \textbf{Path Planning} \\
    A* Algorithm, \\
    RRT, \\
    Dynamic Programming, \\
    Behavior Planning
};
\node [block, below=of path_planning] (control) {
    \textbf{Control} \\
    Reinforcement Learning, \\
    Model Predictive Control, \\
    Hybrid control combined with PID, Sliding mode control etc\\
    Steering/Throttle/Brake, \\
    Actuator Commands
};

\draw [arrow] (perception) -- (localization);
\draw [arrow] (localization) -- (path_planning);
\draw [arrow] (path_planning) -- (control);

\end{tikzpicture}
\caption{Block diagram of autonomous driving system components.}
\label{fig:autonomous_driving}
\end{figure}
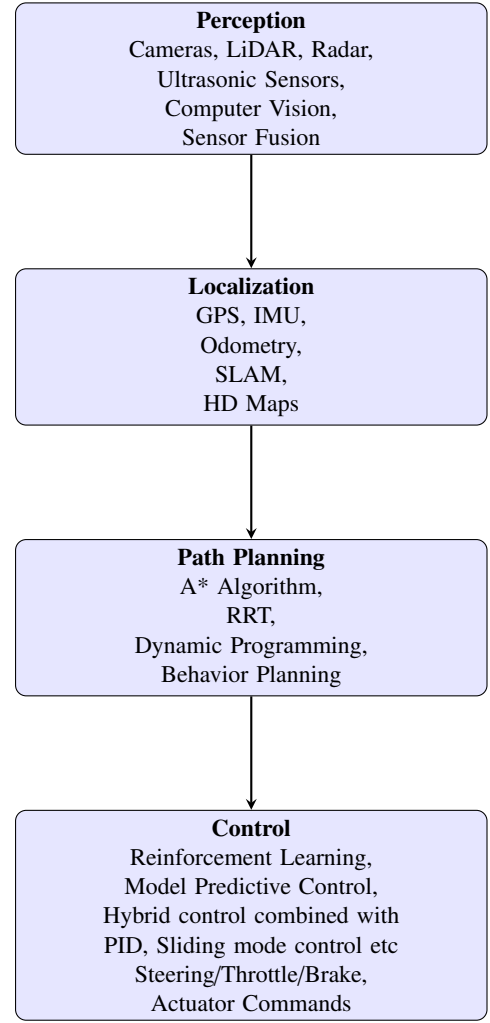

Autonomous driving systems are typically organized into four major functional modules: perception, localization, path planning, and control. 
Fig \ref{fig:autonomous_driving} illustrates the key subsystems of a modern autonomous driving architecture. Among these modules, the control layer plays a critical role in ensuring accurate trajectory tracking and vehicle stability under varying operating conditions. Model-based control strategies, particularly Model Predictive Control (MPC), have been widely adopted in autonomous vehicle research due to their ability to explicitly handle constraints and optimize future control actions over a prediction horizon.

Control methods in general, can be broadly classified into two categories: model based and data driven. \authcite{Benosman}{1234567}
Model based controllers, such as Model Predictive Control (MPC) and Proportional Integral Derivative (PID) control, rely on mathematical representations of vehicle dynamics. These methods offer higher interpretability and predictability. In contrast, data-driven approaches, such as Reinforcement Learning (RL), learn control policies from real world or simulated data, adapting to complex and uncertain environments without requiring fixed models. However, data-driven methods often lack full interpretability and depend heavily on the quality and quantity of training data, which makes provable safety difficult to achieve \authcite{Kiran et al.}{9351818}, \authcite{Zhao et al.}{10682977}. Many AV systems combine these approaches, combining advantages of both approaches. The model based methods provide robustness and the data driven methods provide adaptability. These can be called as hybrid methods.
\\Table \ref{tab:control_algorithms} provides a list and description of the widely researched control algorithms used for Autonomous Vehicles (AVs). Of these, the Adaptive Model Predictive Control (AMPC) and the Reinforcement Learning (RL) methods provide the best outcomes and are widely researched for modern AV control.

The control strategies of commercial autonomous vehicles are not open source and not well documented. However, some technical blogs provide some pointers. The blog post titled "Tesla’s End-to-End Deep Learning Approach" on  \authcite{Think Autonomous}{thinkautonomous2024tesla} talks about Tesla's transition from a modular autonomous driving architecture to a fully end-to-end deep learning system with the release of FSD v12. \authcite{Lu et al.}{https://doi.org/10.48550/arxiv.2212.11419}, \authcite{Bronstein et al.}{https://doi.org/10.48550/arxiv.2210.09539} and \authcite{Gulino et al.}{waymax} are white papers from Waymo. These papers demonstrate Hierarchical Model-Based Imitation Learning for Planning in Autonomous Driving, Imitation with Reinforcement Learning, and Waymax, an Accelerated, Data-Driven Simulator for Large-Scale Autonomous Driving Research.
Despite advancements, control systems face challenges in edge cases, such as adverse weather or unpredictable human behavior, which can degrade sensor accuracy and decision-making \authcite{Chougule et al.}{10335609}. For instance, LiDAR struggles in heavy precipitation, and vision-based systems may misinterpret ambiguous road signs. These limitations, coupled with potential over-reliance on automation, underscore the need for robust control methods. MPC, particularly its adaptive variants, dominates research due to its ability to handle constraints and dynamic environments, as evidenced by its prevalence in recent studies \authcite{Stano et al.}{STANO2023194}. 
\\This review explores traditional MPC, Adaptive MPC, and hybrid methods, analyzing their advantages and shortcomings for path tracking and lane following in AVs. The primary objective of this paper is to provide a structured review of Adaptive Model Predictive Control methods for autonomous vehicles and to consolidate key adaptation mechanisms reported in the literature. In addition to the review, a demonstrative implementation is presented as an illustrative case study to show how curvature- and speed-based adaptation mechanisms can be integrated within a nonlinear MPC framework. The implementation is intended to support understanding of practical design considerations rather than to propose a new control algorithm.
\\
\\
The main contributions of this paper are:
\begin{itemize}
\item A structured review of Adaptive MPC methods for autonomous vehicles.
\item A comparative discussion of adaptation mechanisms and their practical implications.
\item A demonstrative simulation study illustrating curvature and speed adaptive MPC in trajectory tracking.
\end{itemize}

\begin{table*}[t]
\caption{Leading Control Algorithms for Autonomous Vehicles in Research}
\label{tab:control_algorithms}
\centering
\small
\renewcommand{\arraystretch}{1.1}
\setlength{\tabcolsep}{5pt}

\begin{tabularx}{\textwidth}{p{2.5cm} >{\raggedright\arraybackslash}X p{3cm}}
\toprule
\textbf{Control Algorithm} & \textbf{Description} & \textbf{References} \\
\midrule

PID Control &
Classical controller using proportional, integral, and derivative terms to minimize trajectory tracking error. Widely used due to simplicity and low computational cost. &
\cite{SHAKOURI201112964} \\
\\
Traditional Model Predictive Control (MPC) &
Optimizes control inputs over a prediction horizon using a vehicle dynamics model while respecting system constraints. Effective for complex and constrained environments. &
\cite{STANO2023194} \\
\\
Sliding Mode Control (SMC) &
Robust nonlinear control technique that handles model uncertainties and disturbances using switching control laws. Often combined with MPC or fuzzy control. &
\cite{9926457,ALIKA2024102120} \\
\\
Fuzzy Logic &
Rule-based controller using fuzzy sets and linguistic variables to handle nonlinear systems and uncertainty. Often integrated with other control methods. &
\cite{ALIKA2024102120} \\
\\
Adaptive MPC&
Enhances MPC by adjusting controller parameters online to account for changing vehicle dynamics and environmental disturbances. &
\cite{9632266, Xie03042025, Yacoub03062025, 8476236, 9963947, 10730094, s23010412, 9701777, KRENER201831, 9453818, Vaskov01022024, 10316752, 10373911, 9536764} \\
\\
Reinforcement Learning Control &
Data-driven control method where optimal policies are learned through interaction with the environment using reward signals. &
\cite{10486336,7830823} \\
\\
Pure Pursuit &
Geometric path-following algorithm that computes steering commands based on a look-ahead point along the reference trajectory. &
\cite{8203557,9193967} \\

\bottomrule
\end{tabularx}

\end{table*}


\section{Scope of this Review}
This review paper focuses on Adaptive Model Predictive Control (AMPC) methods for autonomous vehicles (AVs), emphasizing control strategies that dynamically adapt to uncertainties and changing conditions in real-time. For the purposes of this review, AMPC is defined as a class of Model Predictive Control (MPC) techniques that explicitly incorporate adaptation mechanisms to modify the system model, cost function, constraints, or prediction horizon based on real-time data or operating conditions. These adaptations ensure robust and efficient control for AV tasks such as path tracking, trajectory following, and obstacle avoidance under dynamic environments (e.g., varying road conditions, traffic patterns, or vehicle parameters).

The scope encompasses the following AMPC methods, which are characterized by their explicit adaptation mechanisms:

\begin{itemize}
\item Gain-Scheduled MPC: Adapts by switching precomputed controllers based on scheduling variables (e.g., vehicle speed) to handle different operating regimes 
\item Online Model Estimation MPC: Updates model parameters (e.g., tire friction, vehicle mass) in real-time using estimators like recursive least squares or Kalman filters 
\item Weight-Adaptive MPC: Dynamically tunes cost function weights to balance objectives like tracking accuracy and control effort 
\item Horizon-Adaptive MPC: Adjusts horizon, such that the prediction horizon N becomes a time-varying parameter Nk, updated at each time step based on a scheduling variable or algorithm.
\item Learning-Based MPC: Employs machine learning, such as Radial Basis Function (RBF) neural networks, to adapt system dynamics, either enhancing a nominal model or replacing it entirely in Model-Free Predictive Control (MFPC) 
\end{itemize}

In addition, this paper reviews hybrid control strategies that integrate Model Predictive Control (MPC) with other control methods, including PID control, fuzzy logic control, sliding mode control, and reinforcement learning.

\section{Traditional MPC}
\begin{figure}[ht]
    \centering
    \begin{tikzpicture}[
    block/.style={rectangle, draw, fill=blue!20, text width=5em, text centered, rounded corners, minimum height=3em},
    subblock/.style={rectangle, draw, fill=blue!10, text width=4.5em, text centered, minimum height=2em},
    line/.style={draw, -Stealth},
    node distance=1cm and 1.5cm
]

    \node[block] (ref) {Reference Trajectory,\\(\( x_{\text{ref},k} \))};
    \node[rectangle, draw, fill=blue!20, minimum width=3cm, minimum height=5cm, right=2.3cm of ref, anchor=center] (mpc) {};
    \node at ([yshift=2.2cm]mpc.center) {MPC Controller};

    \node[subblock] (predictor) at ([yshift=1.5cm]mpc.center) {Predictor,\\(\( x_{\text k+1} \))};
    \node[subblock] (optimizer) at ([yshift=0.5cm]mpc.center) {Optimizer,\\min J};
    \node[subblock, below=0.1cm of optimizer] (cost) {Cost Function, \\ J};
    \node[subblock, below=0.1cm of cost] (constraints) {Constraints};


    \node[block, right=0.7cm of mpc] (vehicle) {Vehicle Dynamics};
    \node[block, below= 0.5cm of vehicle] (sensors) {Sensors Perception};

    \path[line] (ref) --node[midway, above] {error} (mpc.west);
    \path[line] (mpc.east) -- (vehicle);
    \path[line] (vehicle) -- (sensors);
  \path[line]
  (sensors.south)
  -- ++(0,-1)
  -- node[above] {\( x_{\text k} \)} ++(-6.5,0)
  -- (ref.south);

\end{tikzpicture}
\caption{MPC Block Diagram}
\label{mpc block diagram}
\end{figure}
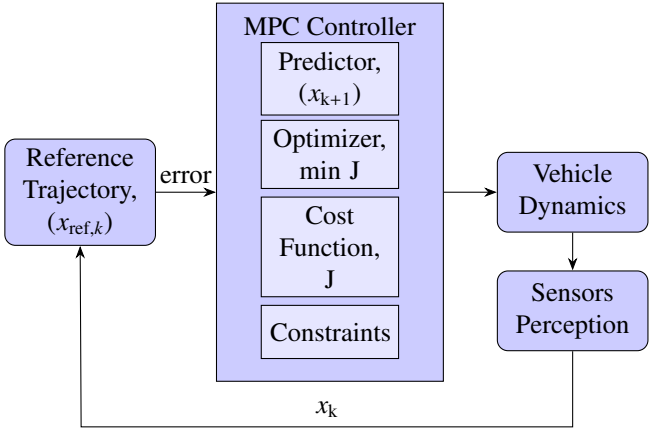

This section will outline the traditional MPC method and its advantages and disadvantages. Model Predictive Control (MPC) is a popular model based approach currently used widely for autonomous vehicles that optimizes control actions over a prediction horizon, considering the vehicle model and constraints. \authcite{Stano et al.}{STANO2023194} gives a detailed review of the model predictive control used in autonomous vehicles. Traditional Model Predictive Control (MPC) is a model based control strategy that optimizes control inputs over a finite prediction horizon, using a vehicle model to consider constraints such as road boundaries and speed limits. It predicts future states and minimizes a cost function to achieve tasks like path tracking. Model predictive control has been explored as a control strategy for autonomous vehicles for more than 2 decades, with early research papers such as \authcite{Raffo et al.}{4768725}, \authcite{Chen et al.}{4787256} and \authcite{Klančar and Škrjanc}{KLANCAR2007460}. Since then MPC is widely researched and several improvements have been explored with Adaptive and Hybrid variants of MPC. The block diagram of traditional MPC for AVs is depicted in Figure \ref{mpc block diagram},  illustrating the interaction of key components.
\\

\subsection{Explanation of Each Block}
\subsubsection{Reference Trajectory Generator}
This block can be thought of the hardware and software that generates the desired trajectory (e.g., position, velocity, yaw) based on navigation goals and environmental data, providing the reference \( x_{\text{ref},k} \) for tracking. 
However, there may be several challenges. Some of these trajectories may be infeasible due to vehicle constraints (e.g., maximum steering angle) or dynamic obstacles, causing tracking errors. Therefore MPC controller has constraints as an input for selecting feasible trajectories.
Even after a feasible trajectory is selected, sudden changes in traffic conditions might make a selected trajectory infeasible , reducing responsiveness. Such abrupt changes can lead to unstable vehicle behavior. In order to overcome these challenges, real time sensor data can be used with adaptive MPC algorithms. Changing vehicle dynamics can be incorporated in the planning. In order to smoothen the dynamically changing trajectories, to avoid abrupt changes, smoothing techniques like splines can be applied. 

\subsubsection{State Estimator (Perception Sensors and Feedback Loop)}
The state estimator can be thought of as combination of Perception Sensors and Feedback to the reference trajectory generator. It is responsible for measuring and estimating the vehicle’s dynamic state variables, such as position, velocity, and yaw rate, by integrating data from onboard sensors, including GPS, IMU, and LiDAR. The error between the measured state \( x_k \) and the reference is fed back to the Model Predictive Controller (MPC) to enable closed loop control. However, noisy sensor measurements can lead to erroneous estimations, which can lead to degraded or even unsafe control performance. In addition, sensor delays, typically in the range of tens to hundreds of milliseconds contribute to delays in response, especially in fast changing driving conditions. This can also lead to unsafe control. Sensor failures, whether transient or permanent, pose a risk to system safety and reliability. These challenges can be mitigated using sensor fusion as well as estimation techniques such as Kalman filtering, which help in reducing noise and improving robustness. 

\subsubsection{MPC Controller}
The MPC controller comprises four major components: the prediction model, the optimizer, and the control function module and the constraints. The prediction model utilizes a simplified dynamic representation of the vehicle to forecast future states over a finite prediction horizon \( N \). 

The evolution of the vehicle state is typically modeled using equations of the type eq\ref{eq:veh_model}

\begin{equation}
\label{eq:veh_model}
\begin{aligned}
\mathbf{x}_{k+1} &= \mathbf{x}_k + v_k \cos(\psi_k)\Delta t \\
\mathbf{y}_{k+1} &= \mathbf{y}_k + v_k \sin(\psi_k)\Delta t \\
\psi_{k+1} &= \psi_k + \frac{v_k}{L}\tan(\delta_k)\Delta t
\end{aligned}
\end{equation}

Here, $\mathbf{x}_k$ and $\mathbf{y}_k$ denote the global position coordinates of the vehicle, $\psi$ is the yaw angle,
$v$ is the longitudinal velocity, $\delta$ is the steering angle, $L$ is the wheelbase,
and $\Delta t$ is the sampling time.
The Model can be linear or Non linear in nature. 
While this model captures basic kinematic behavior, it suffers from model mismatch and fails to represent complex physical phenomena such as tire slip, suspension dynamics, and road irregularities. To overcome these limitations, adaptive modeling approaches such as Online Model Estimation can be employed to update model parameters in real-time. 

The optimizer in MPC solves a constrained optimization problem to compute the control input sequence that minimizes a cost function over the horizon. A typical cost function takes the form eq\ref{eq:cost_fn}
\begin{equation}
\label{eq:cost_fn}
\begin{split}
\min_{u_0, \dots, u_{N-1}} J = \sum_{k=0}^{N-1} \left( \| x_{k+1} - x_{\text{ref},k} \|_Q^2 + \| u_k \|_R^2 \right) + \\ \| x_N - x_{\text{ref},N} \|_P^2,
\end{split}
\end{equation}
subject to the vehicle dynamics \( x_{k+1} = f(x_k, u_k) \) and state and control constraints \( x_k \in \mathcal{X}, u_k \in \mathcal{U} \). The optimization step is computationally demanding, particularly for nonlinear or long-horizon problems, making it difficult to meet real time control deadlines. However since the problem is not convex in nature, the solution can get stuck in a local minima leading to sub optimal solutions. To handle this problem, Methods like interior point or sequential quadratic programming are proposed to solve these problems. Making the model simpler, shortening the prediction horizon, or using techniques like warm starting or rewriting the problem in a simpler (convex) form are some other options.
The MPC controller applies the input from the optimized sequence typically steering, throttle, or braking to the actual vehicle. However, this step must account for actuator constraints and dynamics. Hence, constraints are an important part of the 
MPC controller. Real world actuators may saturate, meaning they cannot execute commands beyond certain physical limits, which can destabilize the vehicle if not properly handled. Furthermore, actuator dynamics, such as delays and nonlinear responses, are often neglected in simplified control models. These limitations can be mitigated by incorporating actuator constraints directly into the optimization problem and using auxiliary controllers or soft constraints to manage the effects of actuator dynamics.

\subsubsection{Vehicle Dynamics}
The vehicle dynamics block represents the actual behavior of the physical vehicle as it responds to control inputs and interacts with its environment. While theoretical models attempt to approximate this behavior, they often omit crucial real-world factors such as tire slip, suspension effects, and drivetrain losses. These unmodeled dynamics can significantly affect vehicle performance, particularly under aggressive maneuvers or on uneven terrain. In addition, external disturbances such as wind gusts, road grade, and surface irregularities introduce uncertainties that are not captured in nominal models. To handle such variability, control techniques like Adaptive MPC can be employed to maintain stability despite bounded disturbances. Disturbance observers or adaptive feedforward control strategies can also be integrated to compensate for known or estimated disturbances in real time.

\subsection{Shortcomings of Traditional MPC as a Standalone Algorithm}
Model based controllers can offer performance guarantees and stability, particularly when using well established control techniques. It is possible to ensure safety critical behavior and incorporate safety constraints in MPC. However, since MPC relies on the vehicle model for prediction, changes in vehicle dynamics can pose a challenge. In addition, MPC is computationally demanding, especially for complex vehicle models and long prediction horizons. Although traditional MPC provides a structured framework for optimal control under constraints, it faces several limitations when applied as a standalone algorithm to autonomous vehicles. One major drawback is model inaccuracy, as fixed linear models fail to capture the time varying nature of vehicle dynamics, such as changing tire friction or varying road conditions. This results in poor tracking performance, particularly during aggressive or emergency maneuvers. Another significant challenge is the computational demand involved in solving optimization problems in real time, particularly for nonlinear systems or extended prediction horizons. Furthermore, traditional MPC does not inherently account for disturbances such as wind, road slope, or sensor uncertainty, which reduces its robustness in unstructured or dynamic environments. Additionally, performance can degrade due to sensor noise and delays, making accurate state estimation essential for reliable control. Finally, actuator limits and their dynamic characteristics are often simplified or omitted in the control formulation, which may result in instability or unsafe behavior. These limitations highlight the need for more advanced approaches such as adaptive, learning based, or gain scheduled MPC methods that can better handle the complex and time varying nature of real world driving conditions.

\subsection{Linear vs. Nonlinear Models}
Linear MPC uses simplified models (e.g., kinematic bicycle model), reducing computational load but sacrificing accuracy. Nonlinear MPC (NMPC) offers superior performance in dynamic scenarios but demands more computational resources and robust solvers. For AVs, NMPC is preferred for high speed or unstructured environments, while linear MPC suits simpler tasks. Adaptive form of MPC can use either Linear or Nonlinear Models. Hence this review does not specifically dedicate separate sections for these two.  

\section{Adaptive MPC }
Variations of traditional MPC like adaptive MPC can be used to handle dynamic and uncertain environments like hilly terrains.
Traditional Model Predictive Control (MPC) uses a fixed model and cost function, while Adaptive MPC updates its system model and/or cost function weights online during operation, allowing it to better handle changing dynamics and uncertainties. Adaptive MPC typically achieves better performance, especially when dealing with nonlinear or time varying systems. 
\\Adaptive MPC can be categorized based on how the "adaptation" is achieved. This section outlines AMPC methods like Gain Scheduled, Online Model Estimation, Learning-Based (including Model-Free), Weight-Adaptive, and Horizon Adaptive MPC. This section excludes Nonlinear MPC unless adaptive.

\begin{table*}[H]
\caption{Summary of Adaptive and Hybrid Model Predictive Control Types for Autonomous Vehicles}
\label{tab:ampc_hybrid_mpc}
\centering
\begin{tabular}{p{2.5cm}p{3cm}p{3cm}p{3cm}p{3cm}}
\toprule
\textbf{Type} & \textbf{Description} & \textbf{Adaptation Mechanism} & \textbf{Advantages} & \textbf{Shortcomings} \\
\midrule
\multicolumn{5}{c}{\textbf{Adaptive Model Predictive Control (AMPC)}} \\
\midrule
Gain-Scheduled MPC & Switches precomputed controllers based on operating conditions (e.g., vehicle speed, road curvature) . & Uses scheduling variables to select controllers for different regimes. & Efficient for predefined conditions; reduces computational load. & Limited to predefined regimes; poor adaptability to unforeseen changes. \\
\midrule
Online Model Estimation MPC & Updates model parameters (e.g., tire friction, vehicle mass) in real-time using estimators  & Employs recursive least squares or Kalman filters to adjust model parameters. & Handles dynamic changes (e.g., road friction); improves tracking accuracy. & Increased computational complexity; relies on estimator accuracy. \\
\midrule
Weight-Adaptive MPC & Dynamically tunes cost function weights based on conditions (e.g., speed, curvature)  & Adjusts weights \( Q_k, R_k, P_k \) using optimization or ML (e.g., PSO-BP neural network). & Balances tracking accuracy and control effort; adapts to varying priorities. & Requires tuning or training; may not handle extreme uncertainties. \\
\midrule
Horizon-Adaptive MPC & Adjusts prediction horizon \( N_k \) based on conditions (e.g., urban vs. highway driving). & Updates \( N_k \) using scheduling variables or algorithms (e.g., based on tracking error). & Optimizes performance for different scenarios (e.g., short horizons for urban driving). & Complex to tune; computational cost varies with horizon length. \\
\midrule
Learning-Based MPC (Model-Enhanced) & Augments system dynamics with ML models (e.g., neural networks) to predict unmodeled dynamics . & ML model \( g_{ML} \) updates parameters \( \phi_k \) via training on real-time data. & Captures complex dynamics; improves prediction accuracy. & Requires extensive training data; computational overhead. \\
\midrule
Learning-Based MPC (Model-Free) & Replaces system model with a data-driven predictor (e.g., neural network). & Neural network \( h_{ML} \) predicts states from past data, updating \( \phi_k \). & Handles nonlinearities and uncertainties; suitable for complex tasks. & Lacks Interpretability; safety not provable; data-dependent performance. \\
\midrule
\multicolumn{5}{c}{\textbf{Hybrid Model Predictive Control}} \\
\midrule
PID-MPC Hybrid & Combines MPC for trajectory tracking with PID for low-level actuator control . & MPC sets optimal setpoints; PID adapts to actuator dynamics. & Reduces computational load; fast actuator response. & PID gains may not adapt to varying conditions; limited robustness. \\
\midrule
Fuzzy Logic-MPC Hybrid & Uses fuzzy logic to adapt MPC parameters (e.g., weights, horizons) based on conditions . & Fuzzy rules adjust parameters based on inputs like curvature, speed. & Enhances adaptability; reduces tracking errors in dynamic scenarios. & Requires extensive rule tuning; limited to rule-based logic. \\
\midrule
Sliding Mode Control-MPC Hybrid & Integrates SMC for robustness with MPC for optimization . & SMC ensures stability under uncertainties; MPC optimizes trajectory. & Robust to disturbances; suitable for unstructured terrains. & SMC chattering causes wear; requires adaptation for AMPC compliance. \\
\midrule
Reinforcement Learning-MPC Hybrid & Uses RL to learn optimal MPC parameters (e.g., cost weights). & RL policy \( \pi \) updates weights \( Q_k, R_k \) based on rewards. & Adapts to dynamic environments; improves tracking performance. & High training complexity; lacks interpretability; safety concerns. \\
\bottomrule
\end{tabular}
\end{table*}


\subsection{MPC Formulation}

Model Predictive Control (MPC) computes an optimal control sequence by solving
the following finite-horizon optimization problem:

\[\begin{split}
\min_{u_0,\dots,u_{N-1}} J =\sum_{k=0}^{N-1}
\left(\| x_{k+1} - x_{\text{ref},k} \|_Q^2 +
\| u_k \|_R^2\right)+ \\\| x_N - x_{\text{ref},N} \|_P^2
\end{split}\]

subject to

\begin{align*}
x_{k+1} = f(x_k, u_k),
x_k \in \mathcal{X},
u_k \in \mathcal{U},
x_0 = x(0)
\end{align*}

where

\begin{itemize}
\item $x_k$ denotes the system state at time step $k$ (e.g., vehicle position, velocity, and yaw angle).
\item $u_k$ denotes the control input at time step $k$ (e.g., steering angle and acceleration).
\item $x_{\text{ref},k}$ is the reference state corresponding to the desired trajectory.
\item $f(\cdot)$ represents the system dynamics model, which may be linear or nonlinear.
\item $Q$ and $R$ are weighting matrices that penalize state tracking error and control effort.
\item $P$ is the terminal cost matrix penalizing the final state deviation.
\item $\mathcal{X}$ and $\mathcal{U}$ denote State constraints (e.g., road boundaries, speed limits) and Input constraints (e.g., steering angle limits, acceleration bounds) respectively.
\end{itemize}

The prediction horizon is denoted by $N$. At each time step $t$, the current
state $x(t)$ is measured and the optimization problem is solved to obtain the
optimal control sequence $\{u_0, \dots, u_{N-1}\}$. Only the first control input
$u_0$ is applied to the system, after which the horizon is shifted forward and
the procedure is repeated.

\section{AMPC Methods}
Each AMPC method modifies the standard MPC formulation to achieve adaptability, addressing uncertainties or changing conditions in AVs (e.g., road friction, vehicle load, or traffic). 
\\
\subsection{Gain-Scheduled MPC}
 Uses a set of precomputed controllers, each designed for specific operating conditions (e.g., speed ranges, road types), selected via a scheduling variable (e.g., vehicle speed $ v $, yaw rate).
Switches precomputed controllers using a scheduling variable \( \sigma(t) \) (e.g., speed):
\\The MPC problem remains the same, but the controller parameters (e.g., model $ f $, weights $ Q, R $, or control law) are selected from a precomputed set based on a scheduling variable $ \sigma(t) $ (e.g., $ \sigma(t) = v(t) $), as shown in eq\ref{eq:gain_scheduled}

\begin{equation}
\label{eq:gain_scheduled}
\begin{split}
J = \sum_{k=0}^{N-1}
\left(
\| x_{k+1} - x_{\text{ref},k} \|_{Q_{\sigma(t)}}^2
+
\| u_k \|_{R_{\sigma(t)}}^2
\right)
+
\\ \| x_N - x_{\text{ref},N} \|_{P_{\sigma(t)}}^2
\end{split}
\end{equation}

\begin{align*}
x_{k+1} = f_{\sigma(t)}(x_k, u_k)
\end{align*}

A lookup table or function maps $ \sigma(t) $ to the appropriate model $ f_{\sigma(t)} $ and weights $ Q_{\sigma(t)}, R_{\sigma(t)}, P_{\sigma(t)} $.
\\This method can be used in AV lane keeping for speed based switching. For example, controllers might be precomputed for speed ranges (e.g., 0-30 km/h, 30-60 km/h), with $ \sigma(t) = v(t) $ selecting the controller.
\\ \authcite{Xie et al.}{Xie03042025} proposes a Gain Scheduling Robust Model Predictive Control (GS-RMPC) approach for the path following control of autonomous independent drive electric vehicles (AIDEVs). The method addresses time varying system dynamics and modeling uncertainties by integrating gain-scheduling mechanisms within a robust MPC framework. Both polytopic and norm-bounded uncertainty representations are considered to enhance robustness. This approach combines offline and online optimization to improve computational efficiency while maintaining control performance.
Authors in  \authcite{Yacoub et al.}{Yacoub03062025} propose a Gain Scheduled Model Predictive Control (MPC) framework for vehicle following applications. Here a single MPC controller is modified to operate under three different driving modes : Speed control, headway control, and emergency braking. The proposed approach allows explicit tuning of two main parameters, namely the standstill distance between the ego and lead vehicles and the desired time headway. The controller is experimentally validated on a passenger vehicle and demonstrates improved smoothness in longitudinal motion by mitigating repetitive acceleration and braking in congested traffic scenarios. Real time data obtained from LiDAR, onboard cameras, and vehicle mounted sensors are used to construct a range-rate representation, from which the desired velocity trajectory for the lower level speed controller is generated.
\\Gain scheduled MPC (GS MPC ) is rarely used in modern AVs as a standalone control algorithm for path tracking due to the complex nature of the problem in real world. While GS MPC is an effective Adaptive Model Predictive Control (AMPC) method that adapts by switching precomputed controllers based on scheduling variables (e.g., vehicle speed, road curvature), its practical challenges make it less suitable as a standalone solution compared to other AMPC methods. AV path tracking requires handling continuous, unpredictable changes (e.g., sudden obstacles, varying friction). GS MPC’s reliance on predefined operating regimes limits its ability to adapt to conditions outside the scheduled set. Some form of gain scheduling is often used with other online adaptive methods of MPC to achieve good results. 

\subsection{Online Model Estimation MPC}

Online Model Estimation MPC continuously updates model parameters (e.g., vehicle mass or tire friction) in real time using parameter estimators such as Recursive Least Squares (RLS) or Kalman filtering. This allows the system model \( f \) to adapt to changing vehicle dynamics and environmental conditions.

The system model \( f(x_k,u_k,\theta_k) \) is parameterized by a vector of parameters \( \theta_k \). For example,

\begin{align*}
   \theta_k = [m, C_f]^{\top}, 
\end{align*}

where \( m \) denotes the vehicle mass and \( C_f \) represents the tire cornering stiffness.

At each time step, an estimator updates the parameter vector based on observed system data:

\begin{align*}
\theta_{k+1} = \text{Estimator}(\theta_k, x_k, u_k, y_k), \quad 
x_{k+1} = f(x_k, u_k, \theta_k),
\end{align*}

where \( y_k \) denotes the measured output (e.g., vehicle position obtained from GPS).

The resulting MPC optimization problem is formulated as shown in eq\ref{eq:online_est}

\begin{equation}
\label{eq:online_est}
\begin{aligned}
\min_{u_0,\dots,u_{N-1}} \; J &= 
\sum_{k=0}^{N-1} \left( 
\|x_{k+1} - x_{ref,k}\|_Q^2 + \|u_k\|_R^2 
\right) \\
&\quad + \|x_N - x_{ref,N}\|_P^2
\end{aligned}
\end{equation}

subject to

\begin{align*}
x_{k+1} = f(x_k, u_k, \theta_k), \quad x_k \in \mathcal{X}, \quad u_k \in \mathcal{U}.
\end{align*}

For example, RLS might estimate $ C_f $ from lateral acceleration and yaw rate measurements, updating the bicycle model:
$$\dot{\psi}_{k+1} = \frac{C_f}{m v_k} \delta_k$$
Applications of Online Model Estimation MPC can include adaptation of a wet road by updating tire stiffness in a kinematic model in order to maintain accurate path tracking. 
\\
This involves continuously updating the parameters of the prediction model within the MPC framework as the system operates. This can be done using techniques like Least squares estimation, Moving Horizon Estimation or Kalman Filtering. In  \authcite{Sakhdari and Azad}{8476236}, an adaptive tube-based nonlinear MPC approach is proposed in which uncertain vehicle parameters are estimated online using a recursive least squares (RLS) method. The updated parameter estimates are incorporated into the prediction model, enabling the controller to account for modeling errors while preserving robustness through tube-based constraint tightening. This approach allows the system to maintain stability and performance despite variations in vehicle dynamics and external disturbances.

In  \authcite{Liang et al.}{https://doi.org/10.1049/iet-its.2020.0357}
, a multi model adaptive control strategy is developed to address uncertainty in tire cornering stiffness. Multiple candidate vehicle models are constructed to represent different operating regimes, and a convex combination of these models is used within the MPC framework. The model weights are adjusted online based on system behavior, allowing the controller to adapt smoothly to changes in road tire interaction without requiring explicit parameter identification.

Similarly,  \authcite{Liu et al.}{https://doi.org/10.1049/itr2.12484}
 presents an adaptive MPC framework that explicitly estimates both tire cornering stiffness and road friction coefficients in real time. These estimates are used to update the internal prediction model as well as the associated constraints, such as slip angle limits, to ensure safe and accurate trajectory tracking. The approach further enhances adaptability by incorporating online tuning of control parameters, enabling consistent performance across a wide range of vehicle speeds and driving conditions.

  \authcite{Vo et al.}{9963947} proposes a robust adaptive path tracking control framework based on a predicted interval formulation for autonomous driving under uncertainty. A set membership based recursive estimation scheme is employed to update uncertain model parameters online, enabling adaptive refinement of the predictive model. The resulting interval based representation is incorporated into a robust MPC framework to ensure constraint satisfaction and closed loop stability. The approach achieves reliable tracking performance under significant modeling uncertainties and varying driving conditions, as demonstrated through extensive simulation studies.

\subsection{Weight-Adaptive MPC}
 This method dynamically tunes the cost function weights $ Q $, $ R $, or $ P $ based on operating conditions, often using optimization or ML techniques (e.g., PSO-BP neural network).
 \\The cost function weights are parameterized as functions of a condition or learned model:
\begin{align*}
    Q_k = Q(\zeta_k), \quad R_k = R(\zeta_k), \quad P_k = P(\zeta_k)
\end{align*}
where $ \zeta_k $ is a condition (e.g., vehicle speed, road curvature) or output of an ML model (e.g., neural network).
The MPC problem can be written as shown in eq\ref{eq:weight_adapt}
\begin{equation}
\label{eq:weight_adapt}
\begin{split}\min_{u_0, \dots, u_{N-1}} J = \sum_{k=0}^{N-1} \left( \| x_{k+1} - x_{ref,k} \|_{Q(\zeta_k)}^2 + \| u_k \|_{R(\zeta_k)}^2 \right) + \\ \| x_N - x_{ref,N} \|_{P(\zeta_k)}^2 
\end{split}
\end{equation}
subject to:
\begin{align*}
  x_{k+1} = f(x_k, u_k), \quad x_k \in \mathcal{X}, \quad u_k \in \mathcal{U}  
\end{align*}
 This method balances tracking accuracy and control effort in path tracking or trajectory following, adapting to conditions like high-speed driving. For example adjusting $ Q $ to prioritize tracking accuracy on curved roads.
\authcite{Hu et al.}{10730094} propose a weight-adaptive model predictive control (MPC) strategy that accounts for variations in road curvature, vehicle speed, and road friction conditions. The controller dynamically adjusts the weighting factors in the MPC cost function to enhance tracking performance under diverse driving scenarios. Co-simulation results obtained using CarSim and MATLAB Simulink demonstrate that the proposed approach achieves safe and stable path tracking across a wide range of operating conditions.

\authcite{Tang et al.}{s23010412} develop a weight-adaptive MPC framework based on a particle swarm optimization–backpropagation (PSO-BP) neural network. In this approach, optimal weighting parameters are obtained offline through automated simulations and subsequently used for online adaptation within the MPC structure. Validation using a PreScan–CarSim–Simulink co-simulation environment shows that the proposed strategy improves tracking accuracy and robustness while satisfying real-time implementation requirements.

Similarly, \authcite{Chang and Suqin}{9701777} employ a fuzzy logic based adaptive mechanism to tune the weighting coefficients of the MPC controller. By adjusting the control weights in response to changes in vehicle speed and road curvature, the method enhances trajectory tracking accuracy and improves vehicle stability under varying driving conditions. The results demonstrate that adaptive weight tuning enables effective performance across a wide range of operating scenarios.

\subsection{Horizon-Adaptive MPC}
Adjusts the horizon such that the prediction horizon $ N $ becomes a time-varying parameter $ N_k $, updated at each time step based on a scheduling variable or algorithm. \authcite{Krener}{KRENER201831} is an example of such an algorithm.
$$N_k = h(\sigma_k)$$
where $ \sigma_k $ could be vehicle speed, tracking error, or environmental data (e.g., distance to obstacles).
The modified MPC problem is shown in eq\ref{eq:horizon_adapt}
\begin{equation}
\label{eq:horizon_adapt}
    \begin{split}
        \min_{u_0, \dots, u_{N_k-1}} J = \sum_{k=0}^{N_k-1} \left( \| x_{k+1} - x_{ref,k} \|_Q^2 + \| u_k \|_R^2 \right) + \\ \| x_{N_k} - x_{ref,N_k} \|_P^2
    \end{split}
\end{equation}
subject to:
$$x_{k+1} = f(x_k, u_k), \quad x_k \in \mathcal{X}, \quad u_k \in \mathcal{U}, \quad x_0 = x(0)$$
\\$ N_k $ can be determined by:
\begin{itemize}
\item Rule-Based: $ N_k = N_{\text{short}} $ if speed is greater than threshold, else $ N_{\text{long}} $.
\item Optimization-Based: Minimize a cost function balancing performance and computation.
\item ML-Based: A neural network predicts $ N_k $ based on data (e.g., traffic density, road curvature).
\end{itemize}
Example: For an AV, $ N_k = 5 $ in dense urban traffic (fast response) and $ N_k = 20 $ on highways (smooth tracking).
\authcite{Kim et al.}{9453818} propose a Model Predictive Control (MPC) framework with a time-varying, non-uniform prediction horizon for autonomous vehicle path following and collision avoidance. The approach employs shorter prediction intervals in the near term to enhance tracking performance in high-curvature segments, while longer intervals are used farther ahead to improve obstacle anticipation and avoidance. This adaptive horizon design enables a balanced trade-off between tracking accuracy and computational efficiency, resulting in improved performance without increasing the overall computational burden.

\subsection{Learning-Based MPC}
Uses machine learning (e.g., neural networks, Gaussian processes) to update or represent system dynamics, either enhancing a model-based approach or replacing the model entirely (MFPC). Adaptation occurs through learning from data to improve predictions. we can further divide these into Model-Enhanced and Model-Free MPC as follows
\\ \subsubsection{Model-Enhanced}
A Machine Learning model augments the system dynamics $ f $, predicting unmodeled dynamics or disturbances,
where $ g_{ML} $ is a learned function (e.g., neural network) with parameters $ \phi_k $, updated via training on data: \\
\begin{align*}
    \phi_{k+1} = \text{Update}(\phi_k, x_k, u_k, y_k)
\end{align*}
\\ The MPC model can then be written as eq\ref{eq:model_enh}\\
\begin{equation}
\label{eq:model_enh}x_{k+1} = f(x_k, u_k, \theta_k) + g_{ML}(x_k, u_k; \phi_k)\end{equation}

\authcite{Vaskov et al.}{Vaskov01022024} propose a friction-adaptive stochastic nonlinear model predictive control (SNMPC) framework for autonomous driving under uncertain road conditions. The method employs a Bayesian learning framework to estimate tire road friction characteristics online using measurements from automotive grade sensors. Two complementary estimators are used to capture both the mean behavior and uncertainty of the friction model, enabling probabilistic characterization of the vehicle road interaction. These uncertainty estimates are incorporated directly into the SNMPC formulation, allowing the controller to adjust its behavior in the presence of high uncertainty, such as during aggressive maneuvers or low-friction conditions. The approach can be regarded as a hybrid of online model estimation and learning based MPC, as it combines data driven friction learning with model based predictive control.

Similarly, \authcite{Kim and Kee}{10316752} employ a neural-network–based approach to adapt the weighting matrices of an MPC controller. In this work, a prediction model based on lateral error dynamics is formulated, and the corresponding MPC objective function is optimized using weight matrices estimated through a radial basis function neural network. The neural network is trained to approximate optimal weighting parameters under varying driving conditions, enabling adaptive adjustment of the controller’s behavior. The proposed method is validated through simulation studies using a CarMaker–MATLAB/Simulink co-simulation framework, demonstrating improved tracking performance and robustness compared to fixed weight MPC strategies.
\\
\subsubsection{Model Free (MFPC)} A model free MPC Completely replaces the model $ f $ with a data driven predictor $ h_{ML} $. This method can be applied for complex tasks like urban driving or drift control, where ML handles nonlinearities or uncertainties. MFPC uses a neural network to predict vehicle trajectory from past sensor data, adapting to urban traffic without a model.
\begin{align*}
  x_{k+1} = h_{\text{ML}}(x_{k-m:k}, u_{k-m:k}; \phi_k).  
\end{align*}

where $ h_{ML} $ (e.g., a neural network) predicts the next state based on past states and inputs, and $ \phi_k $ is updated via data.
\begin{align*}
    \phi_{k+1} = \text{Update}(\phi_k, x_k, u_k, y_k)
\end{align*}
The MPC problem can be written as in eq\ref{eq:model_free}
\begin{equation}
\label{eq:model_free}
\begin{split}
    \min_{u_0, \dots, u_{N-1}} J = \sum_{k=0}^{N-1} \left( \| x_{k+1} - x_{ref,k} \|_Q^2 + \| u_k \|_R^2 \right) + \\ \| x_N - x_{ref,N} \|_P^2
\end{split} \end{equation}
subject to:
\begin{align*}
    x_{k+1} = h_{ML}(x_{k-m:k}, u_{k-m:k}; \phi_k), \quad x_k \in \mathcal{X}, \quad u_k \in \mathcal{U}
\end{align*}
Example: In MFPC, a neural network trained on sensor data (e.g., GPS, LIDAR) predicts vehicle states for steering control. 
\\
\authcite{Nguyen et al.}{10373911} employ deep neural networks (DNNs) to reduce the computational burden associated with Model Predictive Control (MPC) in adaptive cruise control applications. By leveraging the function approximation capability of deep learning, the proposed approach replaces computationally intensive components of the MPC optimization process, enabling faster online execution while preserving control performance.

Similarly, \authcite{Rokonuzzaman et al.}{9536764} present an MPC framework that integrates a neural network–based vehicle dynamics model learned from data collected under diverse driving conditions. The learned model captures complex nonlinear vehicle behaviors that are difficult to represent using conventional analytical models, thereby improving prediction accuracy over the control horizon. The enhanced prediction capability leads to superior tracking performance, particularly in scenarios involving parametric uncertainties and nonlinear dynamics. Simulation results demonstrate that the proposed learning-based MPC approach outperforms conventional model-based MPC in terms of accuracy and robustness across varying road conditions.

\section{Hybrid Methods}
To enhance robustness, Model Predictive Control (MPC) is often combined with other control algorithms, making use of their complementary strengths while retaining the optimization capability of MPC. These hybrid schemes are closely aligned with the principles of Adaptive MPC (AMPC), as they incorporate adaptation mechanisms to cope with varying operating conditions. The current research in hybrid methods can be broadly categorized into following types.

\subsection{PID and MPC Hybrid}
In the PID-MPC hybrid approach, MPC is responsible for high level trajectory tracking and generates optimal setpoints, while Proportional-Integral-Derivative (PID) controllers handle low-level actuator commands such as throttle and steering. This division of tasks reduces the computational burden on MPC and ensures fast dynamic responses through PID. The hybrid structure is widely used in practice, for instance in Apollo’s control framework for lane keeping, where MPC ensures smooth trajectory planning and PID guarantees effective actuator control. \authcite{Chu et al.}{9715995} present a trajectory planning and tracking framework in which artificial potential fields are employed to generate reference trajectories, while an MPC augmented with PID feedback is used for trajectory tracking. Although effective in improving tracking performance, the reliance on fixed PID gains limits the controller’s adaptability to varying vehicle dynamics and road conditions. This limitation motivates the incorporation of adaptive or gain-scheduled mechanisms within the MPC framework to enhance robustness across diverse operating scenarios.

In a related direction, \authcite{Norlund et al.}{NORLUND202419} propose a control architecture that integrates a conventional PID controller with an MPC-based supervisory layer. In this formulation, the MPC provides a feedforward control action that augments the PID output, allowing a smooth transition between pure PID control and MPC dominated operation. The blending between the two control actions is governed by a single tuning parameter, enabling improved performance and stability across different driving regimes while maintaining computational efficiency.

\subsection{Fuzzy Logic and MPC Hybrid}
The fuzzy logic-MPC hybrid incorporates fuzzy inference to adjust MPC parameters such as prediction horizons and cost function weights according to external conditions like vehicle speed or road curvature. This enhances adaptability in uncertain environments and improves tracking performance in dynamic driving scenarios. Nevertheless, the effectiveness of this approach depends heavily on the quality and completeness of the fuzzy rule base, which requires extensive tuning and may still struggle under extreme uncertainties. For example, in \authcite{Chang and Suqin}{9701777} , \authcite{Zhang et al.}{Zhang2024Adaptive} employ fuzzy logic based adaptation mechanisms to dynamically adjust prediction horizons and weighting parameters in MPC for path following applications. In these works, an adaptive velocity planner generates speed profiles based on road curvature and friction conditions, while the fuzzy MPC controller regulates steering and tire forces to ensure stable and accurate trajectory tracking. By continuously tuning the control parameters in response to changing driving conditions, the proposed approaches enhance robustness and tracking performance, particularly in scenarios involving varying road geometry and surface characteristics.

\subsection{Sliding Mode Control and MPC Hybrid}
The sliding mode control (SMC) MPC hybrid exploits the robustness of SMC against disturbances and model mismatches, while using MPC to optimize reference trajectories. SMC provides inherent stability under uncertainties, making the hybrid approach suitable for challenging scenarios such as unstructured or slippery terrains. However, the chattering effect associated with SMC can cause actuator wear and needs to be mitigated through adaptive schemes to maintain AMPC compliance. Applications of this hybrid strategy have demonstrated improved robustness and stability for autonomous vehicle path tracking in adverse conditions such as icy or low-friction roads. \authcite{Dai and Wang}{s23083844} propose a hybrid control framework that integrates robust sliding mode control (SMC) with tube-based model predictive control (MPC). In this approach, MPC is employed to generate the nominal control input for trajectory tracking, while an auxiliary sliding mode controller is designed to regulate the error dynamics between the actual and nominal systems. The sliding surface and reaching law are formulated to ensure robustness against disturbances and modeling uncertainties, allowing the actual system to closely follow the nominal trajectory generated by the MPC. This combined strategy effectively leverages the constraint-handling capability of MPC and the robustness of SMC.

Similarly, \authcite{Ao et al.}{9529225} and \authcite{El Atwi and Daher}{9665575} investigate hybrid control architectures that integrate model predictive control with super-twisting sliding mode control for autonomous vehicle trajectory tracking. In these works, MPC is used to generate reference control actions, while the super-twisting sliding mode controller compensates for uncertainties and external disturbances, ensuring stable and robust tracking performance under varying operating conditions.

\subsection{Reinforcement Learning with MPC}
While standalone reinforcement learning (RL) based control is excluded from this review, hybrid RL--MPC approaches are highly relevant when they incorporate adaptive MPC mechanisms. RL can complement MPC by learning optimal policies or tuning parameters, thereby improving adaptability and robustness. In such frameworks, RL is used to adjust components of MPC such as cost function weights or model parameters. For example, an RL agent may learn a policy \( \pi \) that dynamically tunes the weighting matrices \( Q_k \) and \( R_k \) in the cost function J, where the policy is updated based on rewards derived from tracking accuracy, stability, or safety performance. 

The integration of RL with MPC offers several advantages. By learning parameter adjustments, RL enhances adaptability in dynamic environments such as dense urban traffic or rapidly changing road conditions. Furthermore, performance improvements can be achieved since the RL agent can optimize MPC parameters either offline during training or online in real time, leading to more precise tracking and better control under uncertainty. However, challenges remain. Training RL agents requires extensive computational resources and large amounts of data, making deployment costly. In addition, RL often suffers from a lack of interpretability, and without carefully designed safety constraints, this can pose risks in safety-critical applications such as autonomous driving. 

Recent studies have demonstrated promising results in this direction. \authcite{Wang et al.}{Wang2023} propose a self-learning framework in which reinforcement learning is employed to adaptively tune the weighting coefficients of a model predictive controller. The method extracts representative features from dynamic traffic scenes and utilizes a risk-aware evaluation mechanism to adjust the MPC cost function accordingly. A risk threshold model is introduced to categorize driving scenarios based on their complexity, which in turn guides the reward design for reinforcement learning. This enables the controller to adapt its behavior to varying traffic conditions and enhances its decision-making capability in complex environments.

Similarly, \authcite{Fischer et al.}{10422605} introduce a safe reinforcement learning–based MPC framework that generates reference trajectories while explicitly accounting for safety constraints. In this approach, reinforcement learning is used to explore control policies beyond the local neighborhood of conventional MPC solutions, enabling improved performance in complex driving scenarios. Safety is enforced through a constrained reinforcement learning formulation, where a handcrafted energy-based safety function defines admissible regions of operation. The integration of learning-based exploration with constraint-aware MPC ensures both improved adaptability and safety in autonomous driving applications.

\section{Implementation of Weight and Speed Adaptive MPC}

This section describes a demonstrative implementation of Adaptive Model Predictive Controller (AMPC), which adapts the cost function weights and target speed according to the curvature and slope of the road. The objective of this implementation is not to introduce a novel control method, but to provide a practical demonstration of adaptive MPC concepts discussed in the preceding sections. The controller is designed as a representative example showing how previewed curvature and slope information can be used to adjust reference speed and cost weights within an NMPC framework. The results serve as an illustrative case study in simulation.
\\The controller is formulated as a nonlinear MPC (NMPC) using a dynamic bicycle model and implemented in CasADi. Two MPC formulations are considered, a Fixed-weight MPC (FMPC) with constant target speed and cost function weights, and an Adaptive MPC (AMPC) that adjusts the target speed and cost function weights based on a preview of the upcoming road. The objective is to compare the performance of FMPC and AMPC on identical trajectories containing both curved and sloped road segments. 
\\Both FMPC and AMPC use identical dynamics, constraints, and solver settings, and are evaluated in the CARLA simulator, on the same trajectories. The AMPC and FMPC formulations share identical vehicle parameters, prediction horizon, discretization scheme, solver configuration, and baseline cost weights as shown in Table~\ref{tab:ampc_parameters} . Performance differences between FMPC and AMPC therefore arise solely from adaptation for target speed and weights based on curvature and slope.

\subsection{Vehicle Model}

The vehicle state is defined as
\[
x_t = \begin{bmatrix} \mathbf{x} & \mathbf{y} & \psi & v_{\mathrm{x}} & v_{\mathrm{y}} & r \end{bmatrix}^\top,
\]
where $(\mathbf{x},\mathbf{y})$ denote the global Cartesian position of the vehicle, $\psi$ is the yaw angle, $(v_{\mathrm{x}},v_{\mathrm{y}})$ are body-frame velocities, and $r$ is the yaw rate. The control input is
\[
u = \begin{bmatrix} a_{\mathrm{x}} & \delta \end{bmatrix}^\top,
\]
with longitudinal acceleration $a_{\mathrm{x}}$ and front steering angle $\delta$.

The front and rear slip angles are computed as
\begin{align*}
\alpha_f &= \arctan\!\left(\frac{v_{\mathrm{y}} + l_f r}{v_{\mathrm{x}}}\right) - \delta, \\
\alpha_r &= \arctan\!\left(\frac{v_{\mathrm{y}} - l_r r}{v_{\mathrm{x}}}\right),
\end{align*}
and the corresponding lateral tire forces are obtained from a linear tire model:
\begin{align*}
F_{yf} &= -C_f \alpha_f, \qquad
F_{yr} = -C_r \alpha_r.
\end{align*}

where $l_f$ and $l_r$ are the distances from the vehicle center of gravity to the front and rear axles, respectively, and $C_f$ and $C_r$ denote the cornering stiffness coefficients of the front and rear tires.

The continuous-time dynamics $\dot{s}=f(s,u)$ are given by
\begin{align*}
\dot{\mathbf{x}} &= v_{\mathrm{x}} \cos\psi - v_{\mathrm{y}} \sin\psi, \\
\dot{\mathbf{y}} &= v_{\mathrm{x}} \sin\psi + v_{\mathrm{y}} \cos\psi, \\
\dot{\psi} &= r, \\
\dot{v}_{\mathrm{x}} &= a_{\mathrm{x}} - \frac{1}{m}F_{yf}\sin\delta + r v_{\mathrm{y}}, \\
\dot{v}_{\mathrm{y}} &= \frac{1}{m}(F_{yf}\cos\delta + F_{yr}) - r v_{\mathrm{x}}, \\
\dot{r} &= \frac{1}{I_z}(l_f F_{yf}\cos\delta - l_r F_{yr}).
\end{align*}

where $(\mathbf{x},\mathbf{y})$ denote the vehicle position in the inertial frame, $\psi$ is the yaw angle,
$v_{\mathrm{x}}$ and $v_{\mathrm{y}}$ are the longitudinal and lateral velocities expressed in the body-fixed frame,
$r$ is the yaw rate, $\delta$ is the steering angle,
$a_{\mathrm{x}}$ is the longitudinal acceleration input, $m$ is the vehicle mass,
$I_z$ is the yaw moment of inertia, and $F_{yf}, F_{yr}$ are the
front and rear lateral tire forces.

\begin{algorithm}[t]
\caption{Weight and Speed Adaptive MPC}
\label{alg:dynamic-nmpc}
\begin{algorithmic}[1]
\Require Reference trajectory, vehicle parameters, MPC horizon $N$, sampling time $\Delta t$
\State Preprocess trajectory to obtain arc-length $s$, curvature $\kappa(s)$, and slope $\alpha(s)$
\State Build dynamic bicycle model and RK4 integrator
\State Formulate NMPC with states $X\in\mathbb{R}^{6\times(N+1)}$, controls $U\in\mathbb{R}^{2\times N}$
\State Initialize solver, adaptive weights, and reference speed $v_{\mathrm{prev}}$
\While{vehicle not at goal}
    \State Measure current state $x_t =(\mathbf{x},\mathbf{y}, \psi, v_{\mathrm{x}}, v_{\mathrm{y}}, r)$
    \State Compute current arc length position $s_t$
    \If{$s_t$ near end of trajectory} 
    \State stop and exit \EndIf
    \For{$k = 0 \dots N$}
    \State $s_k \gets s_t + v_x \Delta t \, k$ \EndFor
   \State max previewed curvature $\kappa_{\max} \gets \max_k |\kappa(s_k)|$
    \State max previewed slope $\alpha_{\max} \gets \max_k |\alpha(s_k)|$
    \State $v_\kappa \gets \sqrt{\dfrac{a_{\mathrm{lat,max}}}{\kappa_{\mathrm{max}}+\varepsilon}}$

    \State $\gamma_{\alpha} \gets 
    \min\!\left(1,\max\!\left(0.7,1-c_{\alpha}\alpha_{\max}^{deg}\right)\right)$

    \State $v_{\mathrm{allowed}} \gets 
    \min\!\left(v_{max},\max\!\left(v_{min},v_\kappa\gamma_{\alpha}\right)\right)$

    \State $v_{\mathrm{ref}} \gets (1-\alpha_v)v_{\mathrm{prev}} + \alpha_v v_{\mathrm{allowed}}$
    \State $v_{\mathrm{prev}} \gets v_{\mathrm{ref}}$

    \State $Q_y^{raw} \gets Q_{y,\mathrm{base}}(1+\gamma_\kappa \kappa_{\max})$
    \State $Q_\psi^{raw} \gets Q_{\psi,\mathrm{base}}(1+\gamma_\kappa \kappa_{\max})$

    \State $Q_y \gets \min(Q_{y,max},\max(Q_{y,min},Q_y^{raw}))$
    \State $Q_\psi \gets \min(Q_{\psi,max},\max(Q_{\psi,min},Q_\psi^{raw}))$

    \State Apply temporal smoothing to adaptive weights

    \State Construct reference trajectory with constant $v_{\mathrm{ref}}$
    \State Solve NMPC and apply first control input $(a_0,\delta_0)$
\EndWhile
\end{algorithmic}
\end{algorithm}

\begin{figure*}[t]
\centering
\begin{subfigure}[t]{0.49\textwidth}
    \caption{FMPC tracking on slope varying track}
    \centering
    \includegraphics[width=\textwidth]{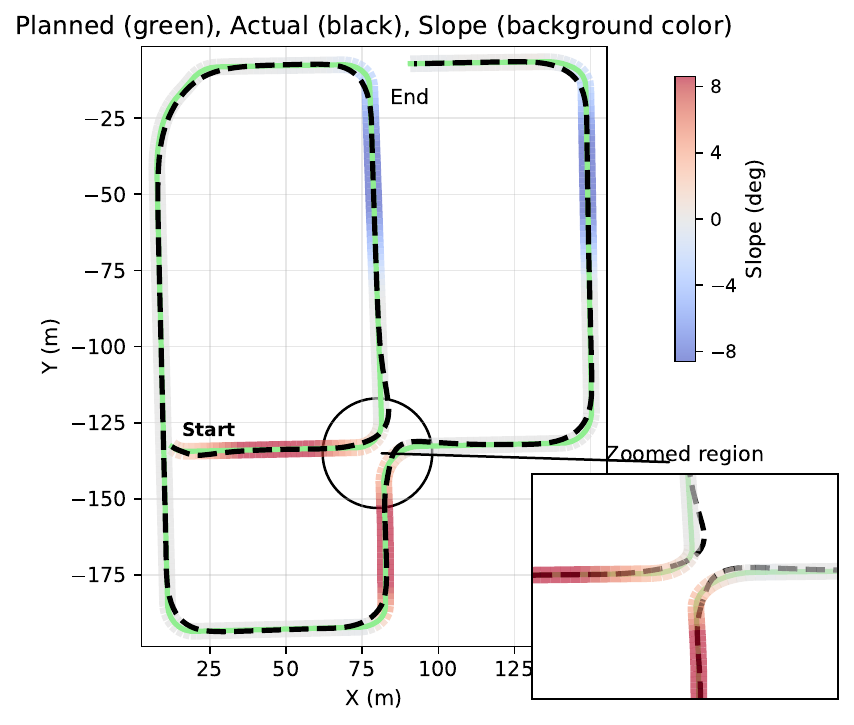}
    \label{fig:fmpc}
\end{subfigure}
\hfill
\begin{subfigure}[t]{0.49\textwidth}
    \caption{AMPC tracking on slope varying track}
    \centering
    \includegraphics[width=\textwidth]{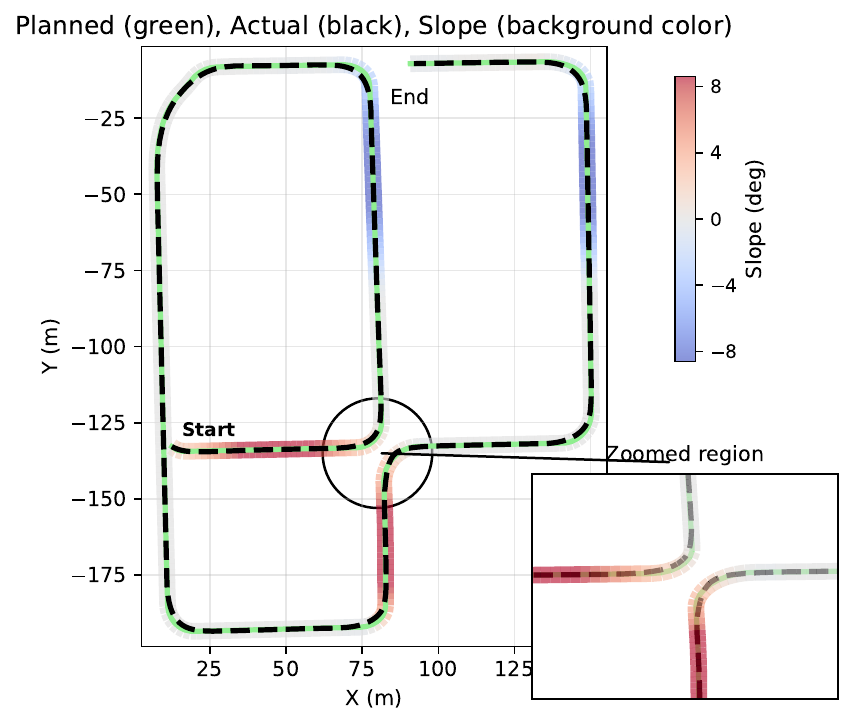}
    \label{fig:ampc}
\end{subfigure}
\caption{Simulation results in CARLA showing planned (green) and actual (black) vehicle trajectories with road slope indicated by the background colormap.}
\label{fig:FMPC_AMPC_slope_compare}
\end{figure*}

\subsubsection*{Discrete time Integration (RK4)} The equations are discretized using an explicit fourth-order Runge--Kutta integrator with fixed time step $\Delta t = 0.10\,\mathrm{s}$ and symbolically implemented in CasADi. The discrete-time state $X_{k+1}$ is computed from $X_k$ and $U_k$ over a time step $dt$: \begin{align*} k_1 &= f(X_k, U_k) \\ k_2 &= f\left(X_k + \frac{dt}{2} k_1, U_k\right) \\ k_3 &= f\left(X_k + \frac{dt}{2} k_2, U_k\right) \\ k_4 &= f\left(X_k + dt k_3, U_k\right) \\ X_{k+1} &= X_k + \frac{dt}{6} (k_1 + 2k_2 + 2k_3 + k_4) \end{align*} While it may be possible to include gravity component in the vehicle model and representing the roll and pitch, the model in this implementation does not use the acceleration due to gravity "g". Adding gravity increases model sophistication and makes the optimization problem more intensive. Including gravitational acceleration due to road grade (g sin $\alpha$) in the prediction model, even when $\alpha$ is estimated from an IMU, provides only reactive compensation that acts after the vehicle has already entered the incline. IMU based slope estimation lacks the critical preview required for anticipative braking before downhill curves or speed reduction ahead of steep descents followed by sharp turns. In contrast, vision based preview of upcoming slope and curvature enables proactive adaptation of reference velocity and stage costs in the MPC. \\This implemented adaptive MPC relies on preview of upcoming road geometry ie, curvature and slope. The controller then lowers the reference velocity and adjusts the cost function weights in advance when it detects an upcoming slope or curvature, preventing the overspeed or undershoot phenomena.

\subsection{NMPC Formulation}

At each control step, the NMPC solves
\begin{align*}
    \min_{u_{0:N-1}}
\sum_{k=0}^{N-1}
\lVert X_{k+1}-X_{\mathrm{ref},k} \rVert_{Q_k}^2
+
\lVert U_k \rVert_{R}^2
+
\lVert X_N - X_{\mathrm{ref},N} \rVert_{P}^2
\end{align*}

subject to the vehicle dynamics, actuator limits, and steering-rate constraints. The nonlinear program is solved using IPOPT. The FMPC employs constant weights and a fixed reference speed, whereas the AMPC updates the reference speed and cost weights online.

\begin{table}[H]
\centering
\caption{Initialization and Adaptation Parameters}
\label{tab:ampc_parameters}
\begin{tabular}{ll}
\hline
\textbf{Parameter} & \textbf{Value} \\
\hline
$Q_{\mathrm{x},{base}}$ & 25 \\
$Q_{\mathrm{y},{base}}$ & 25 \\
$Q_{\psi,{base}}$ & 40 \\
$Q_{v_{\mathrm{x}},{base}}$ & 20 \\
$R_{acc}$ & 0.4 \\
$R_{steer}$ & 15 \\
$R_{\Delta steer}$ & 350 \\
$Q_{y,min}, Q_{y,max}$ & 10,\; 80 \\
$Q_{\psi,min}, Q_{\psi,max}$ & 5,\; 40 \\
$a_{lat,max}$ & 6 m/s$^2$ \\
$c_{\alpha}$ & 0.007 deg$^{-1}$ \\
$v_{min}, v_{max}$ & 8,\; 25 m/s \\
$\alpha_v$ & 0.3 \\
$\alpha_Q$ & 0.20 \\
$\gamma_\kappa$ & 15 \\
\hline
\end{tabular}
\end{table}

\subsection{ Curvature and Slope Adaptive Strategy for AMPC}
The proposed AMPC uses road curvature and slope information extracted from the preview of trajectory to adapt the controller to those particular road segments. Curvature and slope each play different roles in the adaptation.

Road curvature changes both the AMPC cost weights and allowable reference speed. Curvature is computed numerically from the reference path geometry. At each control step, curvature is previewed over the AMPC prediction horizon, and a single effective curvature value is obtained as the maximum absolute curvature within this horizon.

The initialization and adaptation parameters used in the simulation study are summarized in Table~\ref{tab:ampc_parameters}. Baseline state and input weights correspond to nominal straight-road operation and remain fixed for the FMPC case. In the AMPC formulation, lateral and heading tracking weights, as well as the reference speed, are updated online based on previewed road geometry.

\subsubsection*{Curvature and Slope Aware Speed Scheduling}

The allowable reference speed is computed using previewed curvature and slope over a finite lookahead window. The maximum curvature $\kappa_{max}$ and slope $\alpha_{max}$ is obtained as shown in the algorithm \ref{alg:dynamic-nmpc}.

A lateral acceleration limited speed is then obtained as
\begin{align*}
    v_{\kappa} = \sqrt{\frac{a_{lat,max}}{\kappa_{max} + \varepsilon}}
\end{align*}

where $a_{lat,max}$ denotes the maximum allowable lateral acceleration and $\varepsilon$ is a small regularization constant preventing division by zero. 

Slope aware attenuation is subsequently applied using the maximum previewed slope angle $\alpha_{max}$ (expressed in degrees):
\begin{align*}
\gamma_{\alpha} = \max\left(0.7,\; \min\left(1.0,\; 1 - c_{\alpha}\alpha_{\max}^{\mathrm{deg}}\right)\right)
\end{align*}

where $c_{\alpha} = 0.007\,\text{deg}^{-1}$ is a fixed slope sensitivity coefficient. It provides 0.7\% speed reduction per degree of slope and is empirically selected. The attenuation factor is bounded as 
$\gamma_{\alpha} \in [0.7,1]$.
The upper bound ensures that slope does not increase the allowable speed, 
while the lower bound preserves numerical robustness and minimum forward progress. 

The resulting speed is bounded within predefined limits:
\begin{align*}
 v_{allowed}
= \min\!\left(v_{max},\, \max\!\left(v_{min},\, v_{\kappa}\gamma_{\alpha}\right)\right).   
\end{align*}

ensuring feasibility and preventing excessive deceleration. Temporal smoothing is applied to maintain continuity  using a first-order discrete filter:
\begin{align*}
    v_{ref} = (1-\alpha_v)v_{prev} + \alpha_v v_{allowed}.
\end{align*}

Temporal smoothing is introduced to ensure gradual variation of the reference speed and adaptive weights. 
This improves numerical robustness of the NMPC solver, prevents abrupt control transients, 
and enhances feasibility when large preview variations occur.

This anticipative scheduling enables speed reduction prior to high-curvature or downhill segments without explicitly incorporating gravitational terms into the prediction model.

\subsubsection*{Curvature Adaptive Cost Weights}
The state weights $Q_{\mathrm{x},base}$, $Q_{\mathrm{y},base}$, $Q_{\psi, base}$, and $Q_{v_\mathrm{x},base}$ penalize errors in longitudinal position, lateral position, heading angle, and vehicle speed respectively. $R_{acc}$ and $R_{steer}$ penalizes acceleration, and steering change, while $R_{\Delta steer}$ is used to smoothen sudden steering changes. 
The lateral position and heading error weights are adapted using the previewed maximum curvature:
\begin{align*}
Q_y = \max\left(Q_{y,\min},\; \min\left(Q_{y,\max},\; Q_{y,\text{base}}\big(1 + \gamma_\kappa \kappa_{\max}\big)\right)\right)
\end{align*}

\begin{align*}
Q_{\psi} = \max\left(Q_{\psi,\min},\; \min\left(Q_{\psi,\max},\; Q_{\psi,\text{base}}\big(1 + \gamma_\kappa \kappa_{\max}\big)\right)\right)
\end{align*}

where $\gamma_\kappa$ is a fixed curvature sensitivity gain, tuned empirically and $\kappa_{max}$ is recomputed at each control step. First-order temporal smoothing is applied in order to prevent abrupt transients:
\begin{align*}
 Q_k^{sm} = (1-\alpha_Q) Q_{k-1}^{sm} + \alpha_Q Q_k.   
\end{align*}

\subsection{Results and Observations}
Fig. \ref{fig:FMPC_AMPC_slope_compare}  compares FMPC and AMPC performance on a trajectory containing combined curvature and slope variations. Both controllers track the path accurately on flat segments. However, the FMPC maintains a near constant high reference speed (approximately 25~m/s), causing overspeed and lateral deviation in downhill curves.

In comparison, the AMPC previews upcoming curvature and slope, reducing the reference speed in downhill sloped and curved regions. Despite the absence of explicit gravity terms in the prediction model, this anticipative speed and weights adaptation effectively compensates for downhill slopes and curves.

\begin{table}[t]
\caption{Comparison of Fixed MPC and Adaptive MPC Performance}
\label{tab:mpc_comparison}
\centering
\small
\renewcommand{\arraystretch}{1.1}

\begin{tabular}{l c c}
\toprule
\textbf{Metric} & \textbf{Fixed MPC} & \textbf{Adaptive MPC} \\
\midrule
Average speed (m/s)           & 20.075 & 18.214 \\
Minimum speed (m/s)           & 0.000  & 0.000  \\
Maximum speed (m/s)           & 24.250 & 24.169 \\
Average deviation (m)         & 0.806  & 0.597  \\
Maximum deviation (m)         & 3.236  & 1.568  \\
RMS lateral error (m)         & 1.013  & 0.697  \\
95th percentile deviation (m) & 2.148  & 1.236  \\
\bottomrule
\end{tabular}

\end{table}

Table~\ref{tab:mpc_comparison} shows the comparison between Fixed MPC and Adaptive MPC implemented as above on various metrics. The average path deviation is reduced from 0.806~m to 0.597~m (approximately 26\% improvement), while the RMS lateral error decreases from 1.013~m to 0.697~m (approximately 31\% reduction). The maximum deviation is reduced from 3.236~m to 1.568~m, corresponding to an improvement of approximately 52\%, indicating substantially improved robustness in high curvature or dynamically varying segments. 

The average vehicle speed is slightly lower for the Adaptive MPC, as the adaptive mechanism reduces the speed on highly curved roads to maintain stability and tracking accuracy. However, the maximum speed achieved on straight road segments remains nearly identical for both FMPC and AMPC, indicating that the adaptive strategy does not unnecessarily limit performance on straight roads. 

These improvements are a result of the curvature and speed dependent adaptation of MPC weights and velocity, which increases tracking priority and moderates vehicle speed on high curvature segments while allowing higher speeds in low curvature roads. This results in improved tracking accuracy, reduced extreme deviations, and more stable and reliable trajectory following performance.









\section*{Nomenclature}

\noindent
\begin{tabular}{@{}ll@{}}
$\mathrm{x}, \mathrm{y}$ & Global position coordinates (m) \\
$\psi$ & Yaw angle (rad) \\
$v_\mathrm{x}, v_\mathrm{y}$ & Longitudinal and lateral velocities (m/s) \\
$r$ & Yaw rate (rad/s) \\

$a_x$ & Longitudinal acceleration (m/s$^2$) \\
$\delta$ & Steering angle (rad) \\

$l_f, l_r$ & Distance from CG to front and rear axles (m) \\
$m$ & Vehicle mass (kg) \\
$I_z$ & Yaw moment of inertia (kg$\cdot$m$^2$) \\

$C_f, C_r$ & Tire cornering stiffness (N/rad) \\
$\alpha_f, \alpha_r$ & Slip angles (rad) \\
$F_{yf}, F_{yr}$ & Lateral tire forces (N) \\

$\kappa$ & Road curvature (m$^{-1}$) \\
$\kappa_{\max}$ & Maximum curvature (m$^{-1}$) \\
$\alpha$ & Road slope (rad) \\
$\alpha_{\max}$ & Maximum slope (rad) \\

$v_{\text{ref}}$ & Reference velocity (m/s) \\
$v_{\kappa}$ & Curvature-limited velocity (m/s) \\
$v_{\text{allowed}}$ & Allowable velocity (m/s) \\

$\gamma_{\kappa}$ & Curvature adaptation gain \\
$\gamma_{\alpha}$ & Slope attenuation factor \\

$Q, R, P$ & MPC weighting matrices \\
$Q_y, Q_{\psi}$ & Lateral and heading error weights \\

$N$ & Prediction horizon \\
$\Delta t$ & Sampling time (s) \\

$X_k, U_k$ & State and control vectors \\
\end{tabular}

\section{Conclusion}
This paper presented a structured review of Adaptive Model Predictive Control (AMPC) methods for autonomous ground vehicles, with emphasis on explicit adaptation mechanisms applied to the system model, cost function, constraints, or prediction horizon. Adaptive mechanism helps to mitigate the shortcomings of traditional MPC arising from model inaccuracies and dynamic environments. Hybrid approaches such as PID-MPC, Fuzzy-MPC, SMC-MPC and RL-MPC improve
robustness and adaptability. In addition to the review, a demonstrative nonlinear MPC implementation incorporating curvature and slope based preview adaptation was presented. The simulation study shows how adaptation of reference speed and cost weights can improve tracking performance without modifying the underlying vehicle dynamics model. Comparative evaluation against fixed weight MPC showed measurable reductions in lateral tracking error and peak deviation. This highlights the practical value of preview based adaptation for trajectory tracking on road segments with challenging geometries. Future research will focus on reducing computational burdens, improving real time adaptability, and integrating AMPC with advanced perception systems to ensure safe and efficient AV operation in complex scenarios.

\end{document}